\documentclass[final,5p,times,twocolumn]{elsarticle}
\usepackage{graphicx}
\usepackage{dcolumn}
\usepackage{bm}
\usepackage{amssymb}
\usepackage{amsmath}
\usepackage{graphicx}
\usepackage{epsfig}
\usepackage[utf8]{inputenc} 
\usepackage{ucs} 
\usepackage{amsmath} 
\usepackage{array}
\usepackage{float}
\usepackage{color}
\usepackage[usenames,dvipsnames]{xcolor}
\usepackage{epstopdf}
\usepackage{natbib}
\usepackage{graphicx}
\usepackage{lineno}
\usepackage{color}
\usepackage{subcaption}
\usepackage{caption}
\usepackage{hyperref}
\definecolor{red}{rgb}{1.0,0.0,0.0}
\definecolor{blue}{rgb}{0.0,0.0,1}

\newcommand{\bea}{\begin{eqnarray}}
\newcommand{\eea}{\end{eqnarray}}
\newcommand{\nn}{\nonumber}
\newcommand{\etal}{\textit{et al. }}
\newcommand{\kB}{k_{\text{B}}}
\newcommand{\csch}{\text{csch}}
\begin{document}

\title{
A comparative study on heat capacity, magnetization and magnetic susceptibility for a GaAs quantum dot with asymmetric confinement}

\author{J. D. Castaño-Yepes}
\address{Instituto de Ciencias Nucleares, Universidad Nacional Aut\'onoma de M\'exico, M\'exico Distrito Federal, C. P. 04510, M\'exico.}
\author{C. F. Ramirez-Gutierrez}
\address{Posgrado en Ciencia e Ingenier\'ia de Materiales, Centro de F\'isica Aplicada y Tecnolog\'ia Avanzada, Universidad Nacional Aut\'onoma de M\'exico Campus Juriquilla, C.P. 76230, Qro., M\'exico.}
\author{H. Correa-Gallego}\address{Instituto Interdiciplinario de las Ciencias, Universidad del Quind\'io, Armenia, Colombia.}
\author{Edgar A. G\'omez}
\address{
Programa de F\'isica, Universidad del Quind\'io, Armenia, Colombia.}

\begin{abstract}
In this work, thermal and magnetic properties for an electron with cylindrical confinement in presence of external electric and magnetic fields have been investigated. We found that the corresponding time-independent Schr\"odinger equation can be separated into the product of the radially symmetric and an axial equation. Moreover, we have obtained a quasi-exact expression for the energy spectrum of the system in terms of the exact solutions for the radial equation and an approximation up to first order for the axial equation. We have calculated the well-known thermal and magnetic properties as the heat capacity, magnetization and the magnetic susceptibility via the canonical partition function. We found that our results for thermal and magnetic properties differ significantly from results previously obtained by others authors (by Gumber et. al.). Moreover, our results are in agreement with the diamagnetic properties of GaAs.
\end{abstract}

\maketitle

\section{Introduction}
Artificial atoms or semiconductor quantum dots (QD's) have been extensively studied from both theoretical and experimental point of view~\cite{Chakraborty:book}. These quantum systems have many applications such as single electron and photon devices \cite{Kastner:2000, Kafanov:2009, Young:2007, Santori:2001}, diode lasers and nano-antennas \cite{Feng:2010, Curto:2010}, as well as applications in bio-medical sensors and solar cells \cite{Gittard:2011, Semonin:2011}. The QD's are also considered as a possible platform for building solid-state nanodevices with applications in quantum information technologies, since it is now possible with the current state of technology to control the number of electrons in such mesoscopic systems~\cite{DiVincenzo:1998, Burkard:1999, Barenco:1995}. Recently, the importance of thermal and magnetic properties due to electrons confined in parabolic QD's has attracted considerable attention due to its significance in various scientific and technical fields. For example, the effects due to the electron-electron interaction in the energy spectrum~\cite{Maksym:1990, Climente:1992, Pfannkuche:1993} and its electronic structure~\cite{Reimann:2002}. The magnetization effect via Rashba spin-orbit interaction as well as the dia- and paramagnetic effects to the total magnetization in cylindrical QD's~\cite{Avetisyan:2016, Boyacioglu:2012,Boyacioglu:2012-2}. Other theoretical studies have focused on thermodynamical
quantities in anisotropic QDs and its effects on the electronic properties~\cite{Nedelkoski:2014, Madhav:1994}, and lately on electronic states of planar QDs containing a few interacting electrons in an externally applied magnetic field.~\cite{Chakraborty:2016}. From the experimental side, some optoelectronic devices in the presence of electric and magnetic fields have shown important results as intense photoluminescence at low-temperature and electron Raman scattering from quantum wires~\cite{Kash:1986, Temkin:1987, Zhao:2006}, the Giant nonlinear susceptibility and the third-harmonic generation (THG) in QD's~\cite{Brunhes:2000,Khurgin:1988, Wang:2005} and it has motivated the discussion about the importance of external fields into theoretical calculations. For example, Atoyan et al. performed calculations for the case of spinless particle in a 2D cylindrical potential~\cite{Atoyan1, Atoyan2}. Sameer et al. have studied the spectral properties for a 2D parabolic QD in presence of Aharonov-Bohm flux field~\cite{Sameer}. Ghaltaghchyan et al. have investigated the diamagnetic properties of the electron gas in a cylindrical nanolayer~\cite{Ghaltaghchyan}. More recently, a variational method has applied for studying the magnetization and the singlet-triplet transition in the ground state of QD's~\cite{Shaer:2016,Mathew:2013}. 
%
%
Recently, Gumber \etal have considered a 3D cylindrical QD in the presence of external electric and magnetic fields~\cite{Gumber:2015}. Particularly, the authors have found an exact solution for the canonical partition function by assuming a free particle solution for the Schr\"odinger equation along of the axial direction and consequently the thermal and magnetic properties of the system were computed. This assumption should not can affect significantly their results since the typical confinement length in the axial direction is smaller than the radial confinement length in semiconductor nanostructures~\cite{Kleemans:2010}. For example, for a realistic QD the height is approximately $5nm$, which is several times smaller compared to the typical lateral extension of about $30nm$ and consequently, the
quantization energy in a semiconductor QD is also determined by the axial direction. In this paper, we focus on the same physical problem, but taking into account the importance of the boundary conditions along the axial direction and we make a comparative study of the electric and magnetic properties with the previous reported results~\cite{Gumber:2015}. This paper is organized as follows: In Sec.~\ref{Model} the Schr\"odinger equation for a single electron in a cylindrical confinement for the effective mass aproximation is considered. The radial part is solved exactly in terms of Fock-Darwing solutions and an approximation to first order is considered for the axial solution. Thus, the canonical partition function as well as the thermodynamic functions are calculated in terms of a quasi-exact solution. For comparison reasons, the calculations have been divided in two cases: the spinless case and the Zeeman efect contribution. In Sec.~\ref{Results} we compare the canonical partition function based on our approach with the solution calculated by Gumber \etal. In this section, the specific heat a constant volume, the magnetization and the magnetic susceptibility are plotted as a function of temperature and external magnetic field. A discussion about our finding is summarized up in Sec.~\ref{summarized}. 

\section{Theoretical Model}\label{Model}
\subsection{Energy spectrum}
Let us consider that the centre of the base of the cylindrical confinement for the spinless particle is chosen as the origin of the three-dimensional coordinate system. Therefore, the potential energy is given by,
\bea
V(x,y,z)=\left\{\begin{matrix}
\infty &z<0\\ 
\frac{V_0(x^2+y^2)}{2\rho^2}&\,\,\, 0\leq z\leq a,\;\;x,y\leq \rho  \\ 
\infty & z>a 
\end{matrix}\right.
\label{V1}
\eea
where $V_0$ is the depth of parabolic potential. The parameters $\rho$ and $a$ defines the radius and height of the cylinder, respectively. Taking into account that the effective mass approximation is extensively used to describe electronic motion in the presence of slowly varying perturbations, we assume the presence of external electric $F$ and magnetic $B$ fields along the axial direction ($z$-direction). Thus, the Hamiltonian of the system is given by 
\bea
\hat{H}=\frac{\left(\hat{p}-e\hat{A}\right)^2}{2\mu}+V(x,y,z)-eFz,
\label{hamiltonian}
\eea
where $\mu$ is the effective mass particle. In concordance with the orientation of the external fields, we consider the vector potential $\hat{A}$ as the symmetric gauge $\hat{A}=\frac{B}{2}(-y,x,0)$ 
and taking into account the symmetry of the problem, it is straightforward to change to cylindrical coordinates representation, more precisely, the Hamiltonian given by Eq.~(\ref{hamiltonian}) reads, 
\bea
\hat{H}&=&-\frac{\hbar^2}{2\mu}\left(\frac{\partial^2}{\partial r^2}+\frac{1}{r}\frac{\partial}{\partial r}+\frac{1}{r^2}\frac{\partial^2}{\partial\theta^2}\right)+\frac{1}{2}\mu\Omega^2r^2\nn\\
&+&i\frac{eB\hbar}{2\mu}\frac{\partial}{\partial\theta}-\frac{h^2}{2\mu}\frac{\partial^2}{\partial z^2}-eFz.
\eea
\begin{figure}[h!]
\centering
\includegraphics[scale=.46]{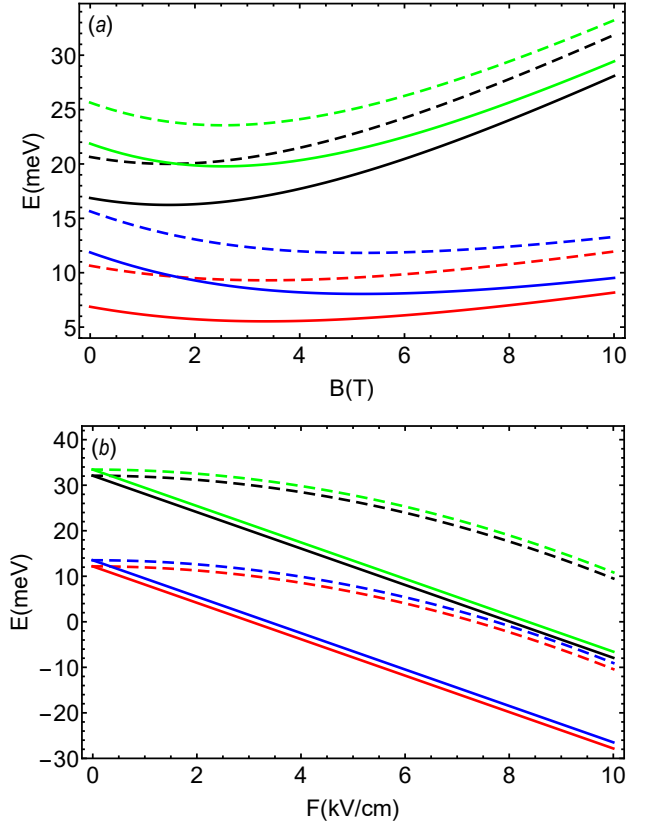}
\caption{(Color online) Discrete energy levels $E_{nlm}$ of the system as a function of the external magnetic field are shown in panel (a). The energy levels $E_{0,1,1}$ (solid red line), $E_{1,1,1}$ (solid black line), $E_{0,2,1}$ (blue solid line) and $E_{1,2,1}$ (solid green line) are computed from the perturbation theory (see Eq.~(\ref{eqfinaleig})) and compared with the results (dashed lines) obtained by Gumber et al. in Ref. \cite{Gumber:2015} from the Eq.~($6$). Similar calculations are shown for the energy levels in panel (b) as a function of the external electric field.  
}\label{EvsBcomparison}
\end{figure}
Thus, the time-independent Schr\"odinger equation is given by
\begin{equation}
\hat{H}\psi(r,\theta,z)=E\psi(r,\theta,z) \end{equation}
where $E$ corresponds to the energy spectrum of the system. To obtain the energy spectrum and wave functions of this system, we can solve partially this Schr\"odinger equation via a separation of variables, it due to that the particle is moving in two dimensions and that the system is symmetric around $z$-direction. Hence after performing a separation of variables with 
\bea
\psi(r,\theta,z)=\mathcal{R}(r,\theta)\xi(z).
\eea
The exact solutions for the radial part are given by
\bea
\mathcal{R}_{n,l}(r,\theta)&=&\sqrt{\frac{b}{2\pi^2l_0^2}\frac{n!}{|l|(n+|l|)!}}\exp\left(-il\theta-\frac{br^2}{4l_0^2}\right)\nn\\
&\times&\left(\frac{\sqrt{br}}{l_0}\right)^{|l|}L_n^{|l|}\left(\frac{br^2}{2l_0^2}\right),
\label{radialfunction}
\eea

\begin{figure}[h!]
\centering
\includegraphics[scale=.44]{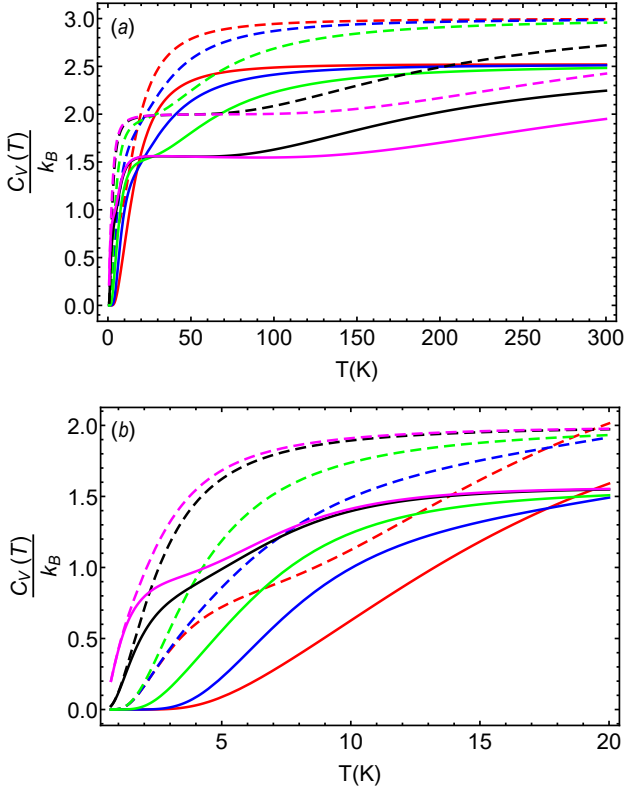}
\caption{(Color online) 
Specific heat $C_{\nu}$ as a function of the temperature is shown in panel (a) for different values of the magnetic field. More precisely, the specific heat is computed from the Eq.~(\ref{cv1storder}) for $B=0T$ (solid red line), $B=5T$ (solid blue line), $B=10T$ (solid green line), $B=30T$ (solid black line) and $B=50T$ (solid magenta line). Numerical results obtained by Gumber et al. in Ref. \cite{Gumber:2015} from the Eq.~($13$) are shown as a dashed lines using the same coding colors for identifying the values of the magnetic field. Panel (b) shows the same comparison for the specific heat, but for a low range of temperatures.}\label{Cvcomp}
\end{figure}
where $\omega_0=\sqrt{V_0/\mu \rho^2}$ is the frequency associated at the confining potential, $\omega_c=eB/\mu$ is the cyclotron frequency, $l_0^2=\hbar c/eB$ is the magnetic length, $\Omega^2=\omega_0^2+\omega_c^2/4$, and $b=\mu\Omega l_0^2/\hbar$. It is worth to mention that these analytical solutions was first found by Fock~\cite{Fock} and later independently by Darwing~\cite{Darwin}. The corresponding eigenvalues are given by
\begin{equation}\label{eqenerradial}
E_{n,l}=(2n+|l|+1)\hbar\Omega-\frac{l\hbar\omega_c}{2},
\end{equation}
with $n=0,1,2,\dots$ and $l=0,\pm1,\pm2,\dots$, which is known Fock-Darwing energy spectrum.  The corresponding axial Schr\"odinger equation for the function $\xi(z)$ reads,
\bea
-\frac{\hbar^2}{2\mu}\frac{d^2\xi(z)}{dz^2}-eFz\xi(z)=E_z\xi(z).\label{eqz}
\eea
In particular, this equation has a general solution 
\bea
\xi(u)=a\text{Ai}(u)+b\text{Bi}(u)
\label{Airy}
\eea
with $u=\left(\frac{2\mu}{eF\hbar^2}\right)(eFz+E_z)$. Moreover $Ai(u)$ and $Bi(u)$ are the Airy functions and $A, B$ are constants that will depend on the boundary conditions. In order to obtain an approximation to the exact solution for the Eq.~(\ref{eqz}) we consider the perturbation theory to first order by assuming that 
\bea
\hat{H}_z=\hat{H}_{z}^{0}+\hat{V},
\eea
where $\hat{H}_{z}^{0}$ defines the free particle Hamiltonian and $\hat{V}=-eFz$ is considered as a perturbation. It is well-known in  basic quantum mechanics that 
\bea
\hat{H}_{z}^{0}| m^{(0)}\rangle=E_m^{(0)}| m^{(0)}\rangle.
\eea
with $\{| m^{(0)}\rangle \}_{m=1,2,3\dots}$ the eigenvectors for free Hamiltonian and its eigenvalues are given by  
\bea
E_m^{(0)}=\frac{\pi^2\hbar^2 m^2}{2\mu a^2}.
\eea
Taking into account the time-independent perturbation theory, it is straightforward  to show that 
the correction to the eigenvalues at first order are given by 
\bea E_m=E_m^{(0)}+E_m^{(1)}=\frac{\pi^2\hbar^2 m^2}{2\mu a^2}-\frac{eFa}{2}.\label{energyfirtsorder}
\eea


The perturbative solution to the energy spectrum associated to the Schr\"odinger equation is given in terms of Eq.~(\ref{eqenerradial}) and  Eq.~(\ref{energyfirtsorder}), it explicitly reads 
\begin{equation}\label{eqfinaleig}
E_{nlm}=(2n+|l|+1)\hbar\Omega-\frac{l\hbar\omega_c}{2}+\frac{\pi^2\hbar^2 m^2}{2\mu a^2}-\frac{eaF}{2}.
\end{equation}
\subsection{Thermal and magnetic properties without spin contribution}
Assuming that the system under consideration is in thermal equilibrium with the much larger bath characterized by the thermodynamic temperature $T$. The fundamental quantity here is the canonical partition function
\begin{equation}\label{cpfdef}
\mathcal{Z}=\text{Tr}\left\{e^{-\beta\hat{H}}\right\}=\sum_{nlm}e^{-\beta E_{nlm}}
\end{equation}
where the summation runs over all possible discrete quantum states, and the parameter $\beta$ defines the inverse temperature, i.e $\beta=1/k_{\text{B}}T$. After replacing the Eq.~(\ref{eqfinaleig}) into the partition function definition, we have that 
\bea\label{eqpart}
\mathcal{Z}&=&\exp\left[\frac{eaF}{2}\beta\right]\sum_{n=0}^{\infty}\exp\left[-(2n+1)\beta\hbar\Omega \right ]\nn\\
&\times&\sum_{l=-\infty}^{\infty}\exp\left[-\beta\left(|l|\hbar\Omega-\frac{\hbar\omega_c}{2}l\right ) \right ]\nn\\
&\times&\sum_{m=1}^{\infty}\exp\left[\frac{\pi^2\hbar^2 m^2}{2\mu a^2}\beta\right],
\eea
The sums on quantum numbers $n$ and $l$ can be performed in an exact form, more precisely by defining the parameters $x=\hbar\Omega$, $y=\hbar\omega_c/2$ and $\alpha=\hbar^2\pi^2/2\mu a^2$, it is
\bea
\mathcal{Z}=\frac{\vartheta_3(0,e^{-\alpha\beta})-1}{8\sinh\left(\frac{x+y}{2}\beta\right)\sinh\left(\frac{x-y}{2}\beta\right)}e^{eFa\beta/2},
\label{partitionfunctionfirtsorder}
\eea
where $\vartheta_3(0,x)$ is the third theta elliptic function \cite{Gradshteyn}.
In the present work, we are interested in the thermodynamic functions as the specific heat at constant volume, magnetization, and magnetic susceptibility of the system. From a theoretical point of view, these physical quantities can be computed via the canonical partition function as follows:
\begin{itemize}
\item Internal Energy $U$:
\bea
U&=&-\frac{\partial}{\partial\beta}\ln Z\nn\\
&=&-\frac{aeF}{2}+\frac{1}{2}\left[(x-y)\coth\left(\frac{x-y}{2}\beta\right)\right.\nn\\
&+&\left.(x+y)\coth\left(\frac{x+y}{2}\beta\right)\right]\nn\\
&-&\frac{\dot{\vartheta}_3(0,e^{-\alpha\beta})}{\vartheta_3(0,e^{-\alpha\beta})-1},
\label{U1storder}
\eea
where the dot represents derivation with respect to $\beta$ parameter. 
\item Specific heat $C_v$:
\bea
C_v&=&-k_{\text{B}}\beta^2\frac{\partial U}{\partial \beta}\nn\\
&=&\frac{\kB\beta^2}{4}(x+y)^2\csch^2\left(\frac{x+y}{2}\beta\right)\nn\\
&+&\frac{\kB\beta^2}{4}(x-y)^2\csch^2\left(\frac{x-y}{2}\beta\right)\nn\\
&-&\frac{\kB\beta^2}{\vartheta_3(0,e^{-\alpha\beta})-1}\left[\dot{\vartheta}_3^2(0,e^{-\alpha\beta})\right.\nn\\
&-&\left.\frac{\ddot{\vartheta}_3(0,e^{-\alpha\beta})}{\vartheta_3(0,e^{-\alpha\beta})-1}\right].
\label{cv1storder}
\eea
\item Magnetization $M$:
\bea
M&=&\kB T\frac{1}{Z}\left(\frac{\partial Z}{\partial B}\right)_T\nn\\
&=&\frac{e\hbar}{2\mu}\frac{\frac{e\hbar B}{2x}\sinh(\beta x)-\mu\sinh(\beta y)}{\cosh(\beta x)-\cosh(\beta y)},
\label{M1storder}
\eea
\end{itemize}
Finally, we mention that it is straigtforward to compute the magnetic susceptibility $\chi$ by considering that $\chi=\frac{\partial M}{\partial B}$. 

\subsection{Thermal properties including the spin contribution}
In order to study the influence of the spin on the thermal properties in a GaAs cylindrical QD, we take advantage of the validity of the effective mass approximation which  simplifies enormously the understanding of various physical phenomena in semiconductors. Thus, we now include spin but do not take into account the spin-orbit interaction and our Hamiltonian is given by
\bea
\hat{H}=\hat{H}_0+\hat{V}+\frac{1}{2}\hbar\omega_cg^{*}\hat{S}_z,
\label{Hconspin}
\eea
where the operator $\hat{S}_z$ denotes the Pauli matrix and  $g^{*}=-0.44$ is the effective Land\'e factor for GaAs quantum dots. It is straightforward to shows that the discrete energy levels for the Hamiltonian given by Eq.~(\ref{Hconspin}) 
\begin{figure}[h!]
\centering
\includegraphics[scale=.46]{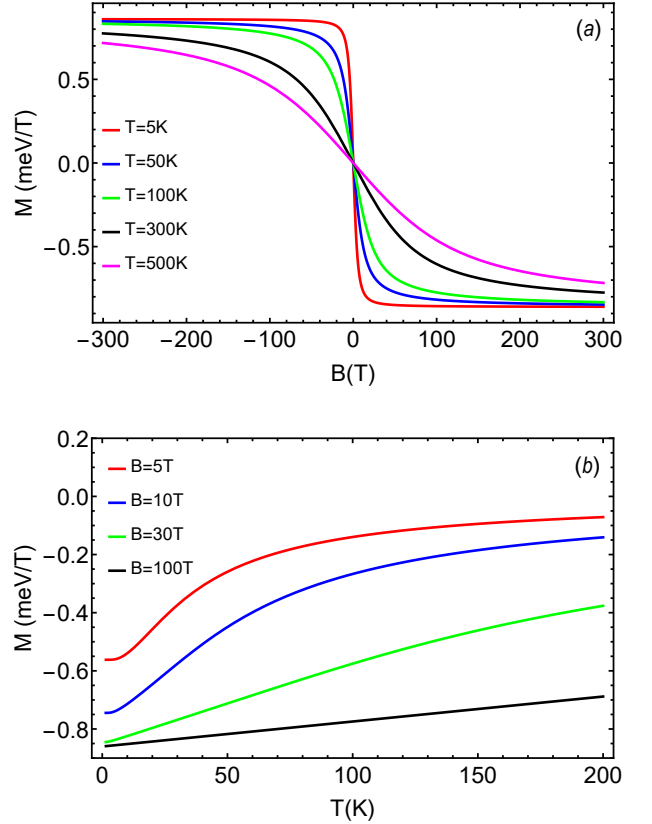}
\caption{(Color online) Magnetization as a function of the external magnetic field computed by the Eq.~(\ref{M1storder}) is shown in panel (a) for a temperature $T=5K$ (solid red line), $T=10K$ (solid blue line), $T=18K$ (solid black line), $T=20K$ (solid magenta line), $T=50K$ (solid green line) and $T=100K$ (solid gray line). Similar calculations are shown in panel (b) for the magnetization but as a function of the temperature at different values of the external magnetic field. For a $B=5T$ (solid red line), $B=10T$ (solid blue line), $B=30T$ (solid green line) and $B=100T$ (solid black line)}\label{M}
\end{figure}

\begin{figure}[h!]
\centering
\includegraphics[scale=.46]{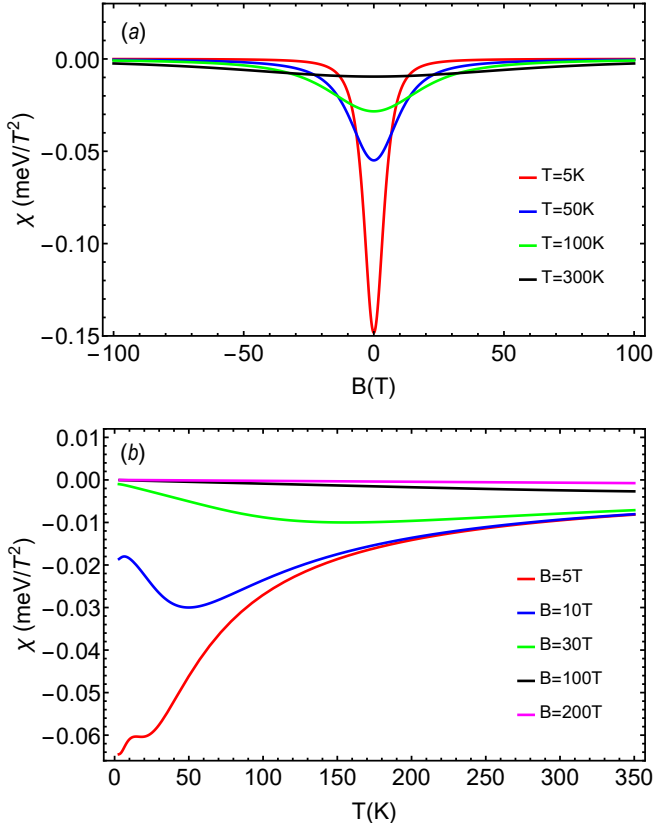}
\caption{(Color online) Magnetic susceptibility as a function of the external magnetic field at different values of the temperature as is shown in panel (a). The results are shown for a temperature $T=5K$ (solid red line), $T=50K$ (solid blue line), $T=100K$ (solid black line) and $T=300K$ (solid black line). Similarly, the magnetic susceptibility as function of the temperature at different values of the magnetic field are shown in panel (b).}\label{chism}
\end{figure}
explicitly reads 
\bea
E_{nlms}&=&(2n+|l|+1)\hbar\Omega-\frac{l\hbar\omega_c}{2}+\frac{\pi^2\hbar^2 m^2}{2\mu a^2}\nn\\
&+&\frac{1}{2}\hbar\omega_cg^{*}s-\frac{eFa}{2},
\label{Econspin}
\eea
since the spin can be described as an additional degree of freedom of the system with the quantum number $s=\pm1/2$. Taking into account the definition into Eq.~(\ref{cpfdef}) it is easy to show that the canonical partition function as well as the specific heat are now given by 
\bea
\mathcal{Z}&=&\frac{\vartheta_3(0,e^{-\alpha\beta})-1}{4\sinh\left(\frac{x+y}{2}\beta\right)\sinh\left(\frac{x-y}{2}\beta\right)}e^{eFa\beta/2}\cosh\left(\frac{g^{*}y}{2}\beta\right)
\label{Zconspin}
\eea
and 
\bea
C_v&=&-k_{\text{B}}\beta^2\frac{\partial U}{\partial \beta}\nn\\
&=&\frac{\kB\beta^2}{4}(x+y)^2\csch^2\left(\frac{x+y}{2}\beta\right)\nn\\
&+&\frac{\kB\beta^2}{4}(x-y)^2\csch^2\left(\frac{x-y}{2}\beta\right)\nn\\
&+&\frac{k_{\text{B}}\beta^2}{4}g^{*2}y^2\text{sech}^2\left(\frac{g^{*}y}{2}\beta\right)\nn\\
&-&\frac{\kB\beta^2}{\vartheta_3(0,e^{-\alpha\beta})-1}\left[\dot{\vartheta}_3^2(0,e^{-\alpha\beta})\right.\nn\\
&-&\left.\frac{\ddot{\vartheta}_3(0,e^{-\alpha\beta})}{\vartheta_3(0,e^{-\alpha\beta})-1}\right].
\label{Cvconspin}
\eea
\section{Results and Discussion}\label{Results}
In this section, we compare our expressions for thermal and magnetic properties produced by the perturbation theory with the corresponding results obtained by Gumber et al. \cite{Gumber:2015} for a GaAs cylindrical QD. In particular, we have considered the parameter values for the system given by $\mu=0.0662m_0$~\cite{Madelung}, $V_0=5$meV, $\rho=1.5$nm, $a=80$nm.   
A comparison for low-lying energy levels of the cylindrical QD is shown in Fig.~\ref{EvsBcomparison}(a). Here the energy levels are computed through the Eq.~(\ref{eqfinaleig}) as a function of the external magnetic field and the numerical results are shown as a solid lines. For comparison purposes, we have plotted the same energy levels of the predicted results by the mentioned authors in the same figure as dashed lines. We have found a significant difference between the results obtained using the perturbative approach when compared with the corresponding solution for the axial Schr\"odinger equation. For example, the energy difference when $F=1$kV is given by 
\begin{equation}
|E^{\text{Gumber}}_{nlm}-E^{(1)}_{nlm}|=\left |\frac{e^2F^2}{2\mu\omega_0^2}-\frac{eFa}{2}\right |\approx 3.77\text{meV}.
\end{equation}
It is worthy of mentioning that this energy shift at the energy levels would be relevant for the experimentalist when they are applying external electric fields along the axial direction of these quantum systems. Similar results for the energy levels when they are plotted as a function of the external electric field are shown in Fig.~\ref{EvsBcomparison}(b). The specific heat $C_{\nu}$ as a function of the temperature is shown in Fig.~\ref{Cvcomp}(a). Here we have computed the specific heat using the Eq.~(\ref{cv1storder}) for different values of the magnetic field. More precisely, we have considered values at $B=0T$ (solid red line), $B=5T$ (solid blue line), $B=10T$ (solid green line), $B=30T$ (solid black line) and $B=50T$ (solid magenta line). We have also plotted the results obtained by Gumber et al. as  dashed lines using the same coding colors for identifying the values of the external magnetic field. We found that a saturation level appears for the specific heat at high values of the temperature and we observe no substantial change in the specific heat. However our results predict a low value for the saturation level as can easily be seen in the figure. It is also compared the specific heat at low temperatures as is shown in Fig.~\ref{Cvcomp}(b) and we found a significant deviation from our results in comparison with the results presented in the reference \cite{Gumber:2015}. For example, if we turn-off the external magnetic field i.e $B=0T$ at temperature approximately of $T=4K$ our model behaves accordingly to the gas of noninteracting particles as is shown as a solid red line in the Fig.~\ref{Cvcomp}(b), but this behavior into the specific heat is misleading in the proposed model by Gumber et al. (see dashed red line) where basically the model suggests interaction between particles. The discrepancy among our results with respect to the calculations presented by Gumber et al. are attributed to the contribution of the third theta elliptic function to the canonical partition function as a consequence of the exact solution to the sums at the Eq.~(\ref{eqpart}). Moreover, it is straigforward to note that the thermal and magnetic properties of the system are independent of the applied electric field in agreement with the previous reported results.\\ 
\begin{figure}[h]
    \centering
    \includegraphics[scale=.42]{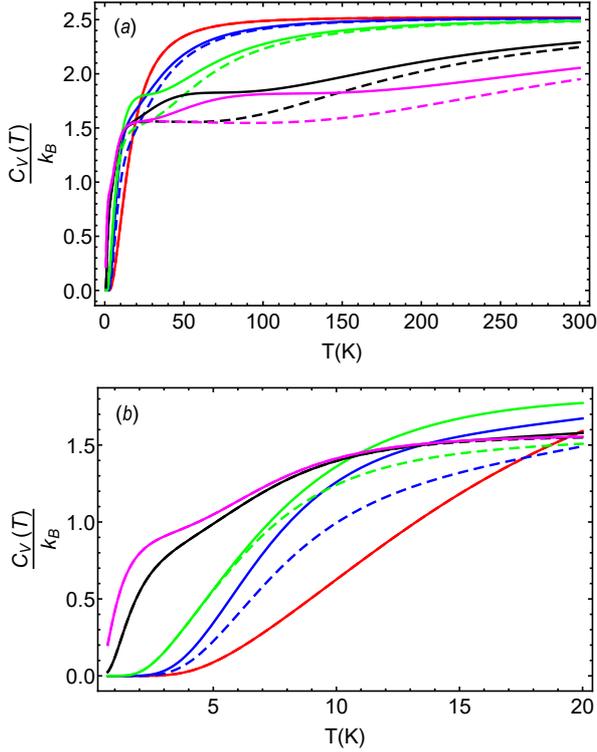}\\
    \caption{(Color line) Comparison between specific heat from Eq.~(\ref{cv1storder}) (dashed lines) and from Eq.~(\ref{Cvconspin}) (continuous lines) as function of temperature at several values of magnetic field. The values for the magnetic field are: 0T (red), 5T (blue), 10T (green), 30T (black), and 50T (magenta).}
    \label{CvTconspin}
\end{figure}

\begin{figure}[h]
    \centering
    \includegraphics[scale=.46]{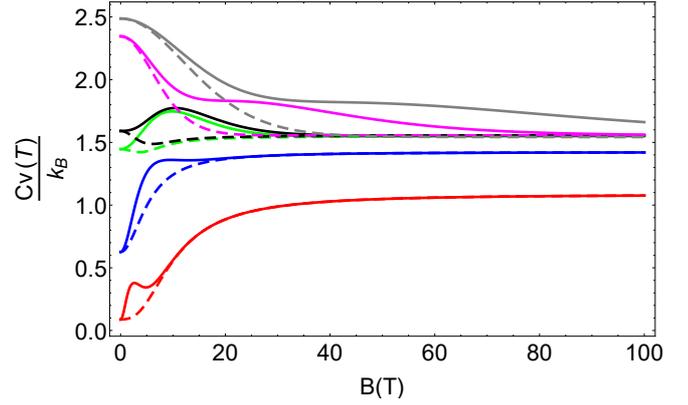}
    \caption{(Color line) Comparison between specific heat from Eq.~(\ref{cv1storder}) (dashed lines) and from Eq.~(\ref{Cvconspin}) (continuous lines) as function of the magnetic field at several temperatures: 5K (red), 10K (blue), 18K (black), 20K (magenta), 50K (green), and 100K (gray).}
    \label{CvBconspin}
\end{figure}
The magnetization of the system as a function of the external magnetic field $B$ is shown in Fig.~\ref{M}(a) at different values of the temperature. Particularly, we found that the magnetization curves are in agreement with a description of non-interacting electrons for the GaAs~\cite{Ingibjorg} since no spin degree of freedom has been considered. Additionally, the magnetization as a function of temperature at different values of the magnetic field is shown in Fig.~\ref{M}(b).\\
It is worthy of mentioning that our approach is consistent with the characteristic behavior of the diamagnetic materials as is well-known for GaAs quantum dots~\cite{Ghaltaghchyan}. More precisely, the magnetic susceptibility $\chi$ takes negative values as 
is shown in Fig.~\ref{chism}(a) and how it is strongly affected by the temperature. We found that the authors in ref.~\cite{Gumber:2015} are erroneously concluding that this system has a paramagnetic behavior.\\

\begin{figure}[h!]
\centering
\includegraphics[scale=.6]{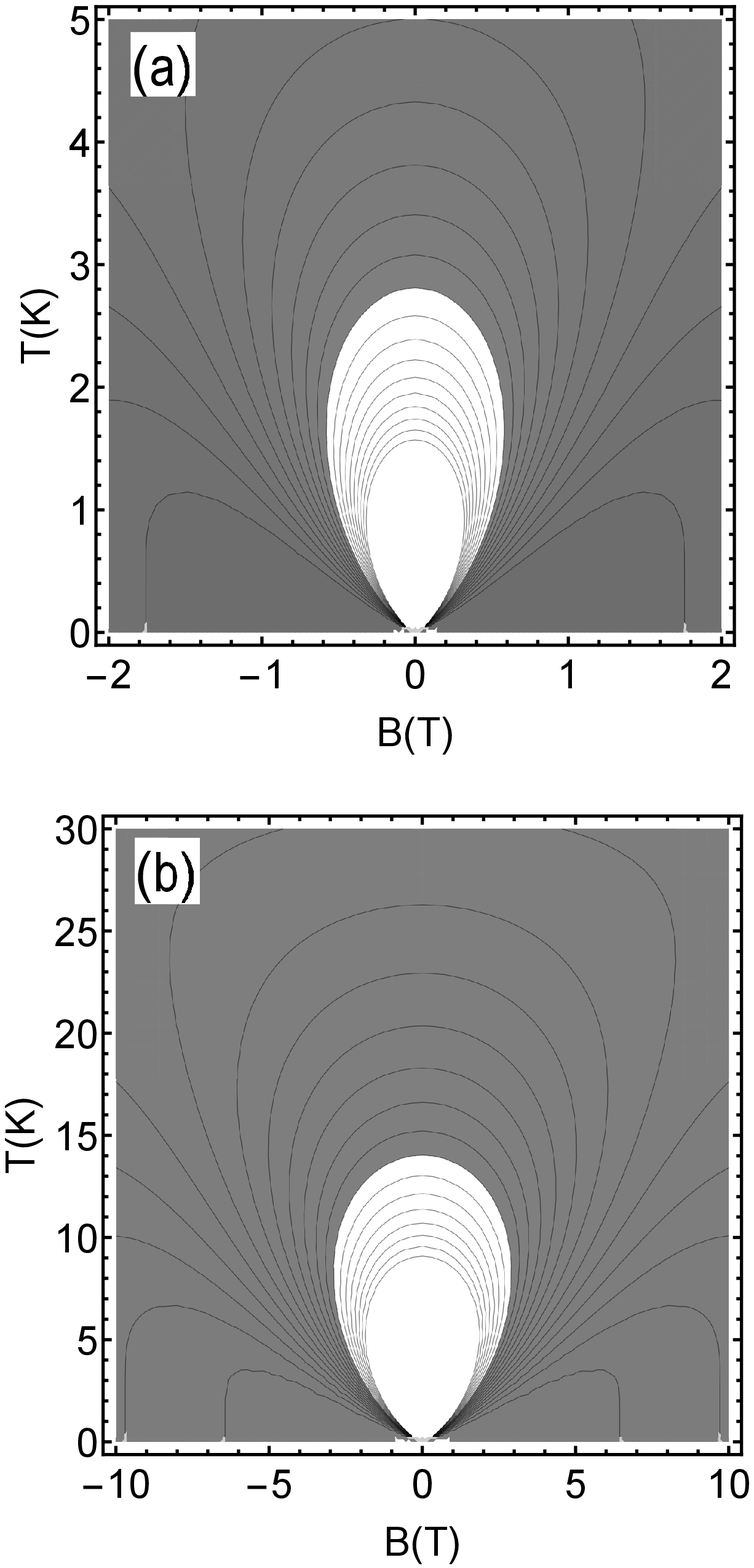}\\
\caption{(Color line) Magnetic phase diagram ofr GaAs as function of $B$ and $T$ for (a) $V_0$=5 meV, and (b) $V_0$=25 meV. The gray region correspond to the diamagnetic phase, and the white region to the paramagnetic phase.}
    \label{phasediagram}
\end{figure}

The spin inclusion shows interesting features in the thermodynamic properties of the system. In Figs. \ref{CvTconspin} (a)-(b), the specific heat is plotted as a function of temperature for several values of the external magnetic field. The dotted lines correspond to the spinless case and the continuous lines when the effective Zeeman effect is added. At high temperature, the value of saturation rises which reflects the new freedom degrees of the model. On the other hand, at low temperatures, the specific heat shows a shoulder structure as a remanent of the Schottky anomaly. Fig. \ref{CvTconspin} (b) indicates that ultra low temperature and high magnetic fields the system absorbs energy in the same way that a spinless system. The behavior of the specific heat with the external magnetic field for different values of temperature is plotted in Fig. \ref{CvBconspin}. The spinless case is also plotted for comparison purposes. For low values of the external magnetic field a peak structure appears and its location changes with the temperature. These peaks are attributed to the Schottky anomaly which is not presented in the spinless configuration. This anomaly is lost at high temperatures.\\
Finally, the magnetic properties of the GaAs with the spin inclusion are given in the phase diagram of Figs. \ref{phasediagram} (a)-(b), as the function of the temperature and the external magnetic field. In these figures, the paramagnetic phase ($\chi(B,T)>0$) correspond to the white region and the diamagnetic phase ($\chi(B,T)\leq0$) at the gray region. It can be noticed that the diamagnetic phase predominates for a large region of values of $B$ and $T$. If the confinement potential is changed as in Fig. \ref{phasediagram}-(b) there are more values of the external parameters in which the system can be in the paramagnetic phase. In both cases, in general, at low temperatures and low values of external magnetic fields, the system is preferred to be in a paramagnetic phase. If the values of $T$ or $B$ increases, the system is diamagnetic. Thus, the actual model is in agreement with previous works, and it can be considered as a good approximation of more elaborated confinement potentials~\cite{Boyacioglu:2012}.

\section{Summary and Conclusions}\label{summarized}

We studied the specific heat, magnetization and magnetic susceptibility for a single spinless particle of in a GaAs quantum dot in the presence of external electric and magnetic fields, by using the perturbation theory at order $\lambda^2$. Our results were compared with the calculation performed by Gumber \etal for the same thermodynamic functions. We found that the expression and curves for both results differ substantially, and we argue that differences come from a wrong treatment of axial part of Schrödinger equation by the mentioned authors, in which they suppose a free particle solution, neglecting the influence of external electric field.
With the presence of electric field at the lower order of approximation, the energy spectrum for both calculations has a difference of energy of 3.77 meV, which makes that the partition function calculated by Gumber \etal have high deviations that propagate error in their thermodynamic functions. The specific heat also was compared, and we found a primary shoulder structure for high magnetic fields and low temperatures, attributed to the interaction of the charged particle with the magnetic field. This structure is present in the Gumber model but for low magnetic fields, even for $B=0$T, which constitute a significant deviation from the expected result. The magnetization, as well as the magnetic susceptibility as a function of temperature and magnetic field, are in agreement with previous calculations for noninteracting electrons in a QD, and the diamagnetic nature of GaAs is in general described. We show that is only necessary the analysis at order $\lambda$. Given that for the magnitude of the electric fields relevant for this study, the eigenvalues of the Hamiltonian are not changed substantially, and therefore, the partition function as well as the thermodynamics functions are well described in the lowest order of the perturbation. Finally, the inclusion of a Zeeman effect allows constructing a para/diamagnetic phase diagram for the GaAs. We found that in general, the GaAs prefer to be in a diamagnetic state, but there exists a region for low temperatures and weak magnetic fields in where the diamagnetic phase appears. This region changes with the intrinsic parameters of the QD (depth and radius).

\section*{Acknowledgments}
The authors J. D. Castaño-Yepes and C. F. Ramirez-Gutierrez acknowledge financial support from the Mexican Consejo Nacional de Ciencia y Tecnolog\'ia (CONACyT). The author E.A.G acknowledges the financial support from Vicerrector\'ia de Investigaciones at Universidad del Quind\'io through research grant No. 752.

\end{document}